\documentclass[acmsmall,screen]{acmart}
\makeatletter
\renewcommand{\@ACM@checkaffil}{}
\newcommand{\paper@authoremails}{%
  \def\email##1##2{\g@addto@macro\@currentaffiliations{, \href{mailto:##2}{##2}}}}
\patchcmd{\@mkauthors@i}
  {\noindent\addresses}
  {\noindent\paper@authoremails\addresses}
  {}
  {\ClassError{acmart}{Could not enable author emails}{Check the author formatting in acmart.cls.}}
\makeatother
\AtBeginDocument{%
  }

\usepackage{graphicx}
\usepackage{tabularx}
\usepackage{threeparttable}
\usepackage{siunitx}
\usepackage{colortbl}
\usepackage{listings}
\usepackage{pifont}
\usepackage{tikz}
\usetikzlibrary{arrows.meta,positioning,calc}
\usepackage{cleveref}
\usepackage{xspace}

\definecolor{listingBackground}{HTML}{F5F7FA}
\definecolor{listingBorder}{HTML}{CAD3DE}
\definecolor{listingKeyword}{HTML}{204E7A}
\definecolor{listingComment}{HTML}{596C60}
\definecolor{listingString}{HTML}{8A4931}
\lstdefinestyle{paper-listing}{
  basicstyle=\ttfamily\footnotesize,
  backgroundcolor=\color{listingBackground},
  rulecolor=\color{listingBorder},
  frame=single,
  framerule=0.35pt,
  framesep=4pt,
  xleftmargin=4pt,
  xrightmargin=4pt,
  aboveskip=5pt,
  belowskip=5pt,
  columns=fullflexible,
  keepspaces=true,
  showstringspaces=false,
  breaklines=true,
  breakatwhitespace=false,
  tabsize=2,
  numbers=none
}
\lstdefinestyle{paper-cpp}{
  style=paper-listing,
  language=C++,
  keywordstyle=\color{listingKeyword}\bfseries,
  commentstyle=\color{listingComment},
  stringstyle=\color{listingString}
}
\lstdefinestyle{paper-trace}{style=paper-listing,language={}}

\graphicspath{{figures/}}

\setcopyright{acmlicensed}
\copyrightyear{2018}
\acmYear{2018}
\acmDOI{XXXXXXX.XXXXXXX}
\acmConference[Conference acronym 'XX]{Make sure to enter the correct
  conference title from your rights confirmation email}{June 03--05,
  2018}{Woodstock, NY}
\acmISBN{978-1-4503-XXXX-X/2018/06}

\newcommand{\sysname}{\textsf{HarnessMinimizer}\xspace}

\definecolor{redlightbg}{HTML}{FCE4EC}   
\definecolor{bluelightbg}{HTML}{E3F2FD}  
\definecolor{figlightbg}{HTML}{E8F5E9}   

\newcommand{\redlight}[1]{%
  \begingroup\setlength{\fboxsep}{1.5pt}%
  \colorbox{redlightbg}{#1}\endgroup}
\newcommand{\bluelight}[1]{%
  \begingroup\setlength{\fboxsep}{1.5pt}%
  \colorbox{bluelightbg}{#1}\endgroup}
\newcommand{\figlight}[1]{%
  \begingroup\setlength{\fboxsep}{1.5pt}%
  \colorbox{figlightbg}{#1}\endgroup}

\newcommand{\secref}[1]{\redlight{\autoref{#1}}}
\newcommand{\tabref}[1]{\bluelight{\autoref{#1}}}
\newcommand{\figref}[1]{\figlight{\autoref{#1}}}

\begin{document}

\title{Efficient Program Reduction for Library Fuzzing Harnesses}

\author{Haiying Li}
\authornote{Authors contributed equally.}
\affiliation{%
  \institution{The University of Hong Kong}}
\email{lihy2022@connect.hku.hk}

\author{Yunlong Lyu}
\authornotemark[1]
\affiliation{%
  \institution{The University of Hong Kong}}
\email{yunlong.lyu97@gmail.com}

\author{Peng Chen}
\affiliation{%
  \institution{Independent Researcher}}
\email{spinpx@gmail.com}

\author{Hao Chen}
\affiliation{%
  \institution{The University of Hong Kong}}
\email{chenho@hku.hk}

\renewcommand{\shortauthors}{Li et al.}

\begin{abstract}
Library fuzzing harnesses translate generated inputs into API calls. When a complex harness triggers a crash, developers must determine whether the cause is a library defect or invalid harness usage. A smaller reproducer can simplify this investigation, but obtaining one requires repeatedly building and executing modified harnesses and analyzing their crashes, making manual or even automatic minimization time-consuming. Moreover, data-dependent statements prevent the reducer to remove unnecessary code and the heavy stack trace symbolization in libraries also slows down the candidate execution. To address these challenges, we present \sysname, a system for efficient minimization of C/C++ library fuzzing harnesses. Its optimized pipeline reuses header analysis and prepared libraries and runtimes to reduce the work required to build and execute each simplified harness during the reduction process. It records and replays input values so that unnecessary input operations can be removed without changing the values supplied to surviving calls. It also uses a fast method to check whether each simplified harness preserves the original crash, avoiding repeated stack trace symbolization. We evaluate \sysname with four reduction engines on 200 crash cases from 21 open source libraries. Compared with the unoptimized pipeline, \sysname achieves speedups of 3.46--4.51$\times$ in average total wall time. Average harness size decreases from 1,352 original tokens to 143--263 remaining tokens after minimization. In crash triage using a large language model, accuracy in identifying harness bugs increases from 28\% with the original harnesses to 44--51\% with the minimized harnesses, while accuracy for library bugs remains at 100\%.

\end{abstract}



\maketitle

\section{Introduction}
\label{sec:introduction}
Library fuzzing searches for bugs by repeatedly exercising library APIs with fuzzer-generated inputs. A fuzzing \emph{harness} connects these inputs to the APIs by constructing arguments, initializing library state, invoking functions, and releasing resources~\cite{manes2021fuzzing,hopper2023}. Developers can write harnesses manually or use tools to generate them. Recent systems use large language models (LLMs) to generate harnesses from library declarations, documentation, and calling relationships~\cite{promptfuzz2024,promefuzz2025,ckgfuzzer2025, fuzzagent2026}, achieving significant improvement over OSS-Fuzz~\cite{oss_fuzz}. For example, PromptFuzz~\cite{promptfuzz2024} utilize LLMs to generate harnesses with fuzzing feedback, achieving 61\% more code coverage than OSS-Fuzz~\cite{oss_fuzz}.

To increase coverage, a harness may support several testing modes, construct complex inputs, and exercise varied API combinations~\cite{promptfuzz2024,graphfuzz2022}. This can require substantial source code to complete. \figref{fig:generated-harness} shows an abbreviated example of a libvpx harness from our evaluation benchmark, whose 433 source lines cover encoder configuration, layer controls, frame construction, and encoding.
On the one hand, the source files for existing fuzzing harnesses are lengthy: libxml2's \texttt{fuzz/api.c}, FFmpeg's \texttt{tools/target\_dec\_fuzzer.c}, and JPEG XL's \texttt{tools/djxl\_fuzzer.cc} contain 3614, 656, and 613 lines, respectively~\cite{libxml2-api-harness,ffmpeg-decoder-harness,libjxl-decoder-harness}. On the other hand, harnesses generated by LLMs share the similar complexity: the 200 harnesses generated by PromptFuzz, PromeFuzz and FuzzAgent has an average length of 273 lines, as shown in \tabref{tab:overall}. Such complexity is not merely an edge case and can substantially complicate the understanding of crashes exposed by a harness.

\begin{figure}[tbp]
  \centering
  \begin{minipage}{\linewidth}
    \footnotesize
\begin{lstlisting}[style=paper-cpp]
132 extern "C" int LLVMFuzzerTestOneInput(const uint8_t *data, size_t size) {
      // ... input validation ...
137   FuzzedDataProvider fdp(data, size);
      // ... mode selection and layer counts ...
157   unsigned int base_width = fdp.ConsumeIntegralInRange<unsigned int>(64, 1920);
      // ... other input values and encoder configuration ...
237   err = vpx_codec_enc_init_ver(&encoder, encoder_iface, &cfg, flags,
                                 VPX_ENCODER_ABI_VERSION);
      // ... initialization checks, layer controls, and image allocation ...
356   int num_frames = fdp.ConsumeIntegralInRange<int>(2, 8);
      // ... frame duration ...
359   for (int frame_num = 0; frame_num < num_frames; ++frame_num) {
        // ... layer updates and frame construction ...
404     err = vpx_codec_encode(&encoder, &img, frame_num * frame_duration,
405                            frame_duration, frame_flags, VPX_DL_REALTIME);
        // ... output retrieval and further controls ...
423   }
      // ... encoder flushing ...
429   vpx_img_free(&img);
430   vpx_codec_destroy(&encoder);
432   return 0;
433 }
\end{lstlisting}
  \end{minipage}
  \caption{Generated libvpx harness, abbreviated from 433 lines. Line numbers refer to the original source.}
  \Description{An excerpt showing input consumption, encoder initialization, repeated frame encoding, and cleanup in a generated libvpx harness. Original line numbers and omission comments identify the selected portions of the source.}
  \label{fig:generated-harness}
\end{figure}

When crashes are discovered during library fuzzing, developers must then determine whether the crash comes from a library defect or invalid API usage in the harness. This \emph{crash triage} depends on understanding the relevant calls and arguments to locate the root cause. However, recognizing such root causes in a large harness can be difficult because the relevant call and its arguments appear among many other API operations. Manually identify the root cause is time-consuming as it requires numerous attempts of simplification in the program code and checks for crash preservation~\cite{creduce2012}.
Program reduction addresses this problem by eliminating code that is irrelevant to the failure, thereby isolating the calls and values necessary to reproduce and explain the crash~\cite{simplifying2002}. 
It automates simplification by repeatedly modifying a program and retaining changes that preserve the behavior being investigated, which largerly reduces the manual effort required to isolate the root cause.

Many program reduction techniques have been developed and evaluated using test cases that trigger compiler bugs. C-Reduce combines transformations specialized for C/C++~\cite{creduce2012}, while Perses uses language grammars to avoid syntactically invalid candidates~\cite{perses2018}. Vulcan introduces additional transformations to expose further simplifications after a reducer reaches a local minimum~\cite{vulcan2023}. WDD uses fragment sizes to guide partitioning and deletion decisions, improving reduction efficiency~\cite{wdd2025}. Their shared focus is on the reduction algorithm itself, improving its efficiency and effectiveness through the design of program transformations and search strategies.

However, directly applying existing program-reduction techniques to harness minimization is prohibitively inefficient. These techniques are largely designed for self-contained failure-inducing programs, such as compiler test cases, where each reduction query primarily transforms and tests a standalone source file. In contrast, reducing a fuzz harness requires repeatedly integrating the modified harness with the target library and its runtime environment, incurring substantial costs from compilation, linking, execution, and crash symbolization, which cannot be ignored. We identify three challenges that existing reduction engines do not address, which make harness minimization expensive and ineffective. First, \textbf{each query repeats costly build and execution preparation}, even though the library headers, target library, and supporting runtimes remain unchanged. Second, \textbf{input-consumption dependencies constrain source deletion}: removing one consumption operation can change the values supplied to surviving library calls even with the same crash input, forcing the reducer to retain otherwise irrelevant code. Third, \textbf{symbolized crash comparison adds diagnostic overhead to each crashing query}, as instruction addresses must be translated into function names and source locations to compare the crash location and calling context with those of the original execution. We elaborate on these three challenges in \secref{sec:challenges}.

We present \emph{HarnessMinimizer}, a system for efficient automatic reduction of C/C++ library fuzzing harnesses. Its key observation is that reduction changes the harness while the library build, headers, and reference crash remain fixed. Building on existing reduction engines, \sysname addresses the three challenges through a shared candidate testing pipeline. It reuses header analysis through precompilation and library code and initialization through a persistent runner, supporting both dynamic and static linking. Call-site replay preserves the recorded input values supplied to surviving operations, allowing the reducer to remove unnecessary consumption calls. A lightweight crash oracle combines diagnostic patterns with stable shared-library offsets, or stack depth as a coarser fallback, to filter candidates without repeated symbolization. After reduction, \sysname embeds the surviving input values into the source and validates the resulting reproducer, exposing the relevant API calls and concrete arguments for debugging.

We evaluate \sysname with four reduction engines---treereduce~\cite{treereduce}, Perses~\cite{perses2018}, WDD~\cite{wdd2025}, and CDD~\cite{cdd2025}---on 200 crash cases from 21 open source libraries, comprising 100 library bugs and 100 harness bugs. Compared with the unoptimized pipeline for each engine, \sysname achieves 3.46--4.51$\times$ speedups in mean total wall-clock time. The original harnesses contain 1,352 source tokens on average; after reduction, the mean remaining size is 143--263 tokens across the four engines. In LLM-based crash triage, minimized harnesses improve accuracy in identifying harness bugs from 28\% to 44--51\%, while accuracy for library bugs remains at 100\%.

This paper makes three contributions:
\begin{itemize}
  \item \textbf{A method tailored to library harness minimization.} We design \sysname to reuse stable build and execution preparation, preserve consumed input values across source deletions, and recognize reference crashes without repeated symbolization, addressing the costs and dependencies of reducing C/C++ library fuzzing harnesses.
  \item \textbf{Demonstrated efficiency and crash triage benefits.} Our evaluation on 200 crash cases from 21 libraries shows 3.46--4.51$\times$ speedups in mean total wall time and a decrease in mean harness size from 1,352 to 143--263 tokens. The minimized harnesses improve LLM accuracy in identifying harness bugs from 28\% to 44--51\%.
  \item \textbf{Integration with multiple reduction engines.} We provide adapters for treereduce, Perses, WDD, and CDD through a shared interestingness-test interface, allowing different reduction strategies to benefit from the same optimized harness testing pipeline.
\end{itemize}

\section{Challenges in Harness Minimization}
\label{sec:challenges}

\begin{figure}[tbp]
  \centering
  \includegraphics[width=\linewidth]{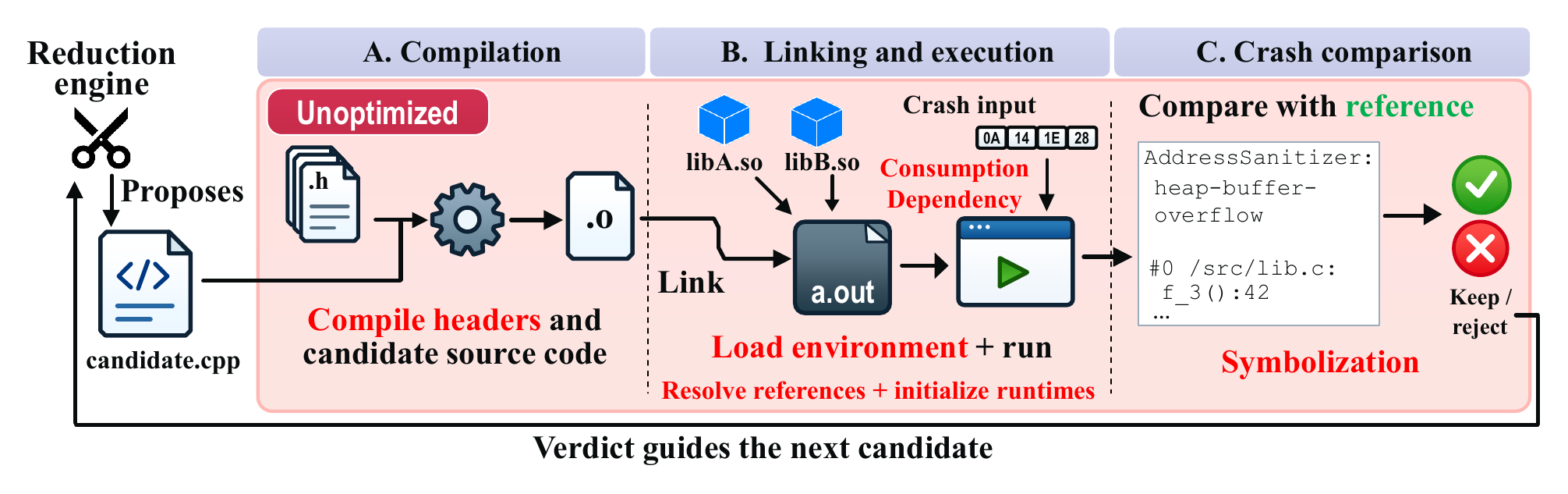}
  \caption{Challenges in existing harness minimization. Testing candidates repeats header processing, linking, runtime initialization, and symbolization, while deleting input operations can change the arguments passed to the library.}
  \Description{A reduction engine proposes a candidate harness, which is compiled with its headers, linked with target libraries, and executed on the crash input. Its symbolized crash report is compared with the reference, and the verdict guides the next candidate. The diagram highlights repeated preparation and diagnostic work, together with input consumption dependencies.}
  \label{fig:challenges}
\end{figure}

A library fuzzing crash is reproduced by a harness, a triggering input, and the library build against which the harness executes. Harness minimization changes the first of these artifacts: it removes or simplifies source code while seeking to retain the crash. A \emph{reduction engine} proposes candidate harnesses, and an \emph{interestingness test} accepts or rejects each candidate. We call one invocation of this test a \emph{query}. However, existing reduction engines only provide different search strategies and source transformations~\cite{creduce2012,perses2018,wdd2025} on the harness code. For minimizing library fuzzing crashes, their effectiveness in this setting also depends on how queries build candidates, supply input values, and recognize the crash. These three responsibilities give rise to the following challenges (\figref{fig:challenges}).

\subsection{Repeated Build and Execution Costs}
\label{sec:challenge-testing}

Each query must turn a modified C/C++ harness into an executable computation. Compilation processes its included headers, checks declarations and types, and generates object code. Library headers can contain substantial amounts of code, including nested includes, inline functions, and templates~\cite{yalla2025}. Removing a few statements from the harness does little to reduce this work: the candidate still needs the same declarations to invoke the remaining APIs. A \emph{precompiled header} (PCH) can cache header analysis~\cite{llvm-pch}, but reusing it requires a stable boundary between the cached declarations and the source being reduced. Template instantiation and code generation can also remain expensive even when header parsing is cached.

Compilation is followed by linking and process startup. With static linking, selected library object code is incorporated into the executable. The linker selects the required objects, resolves function and data references and adjusts addresses to match the executable's layout~\cite{gnu-ld-options}. Compiling the library beforehand does not eliminate this work when each candidate is linked. With dynamic linking, the executable refers to shared library binaries loaded for execution. Starting the executable requires loading dependencies, resolving runtime references, and running initialization code~\cite{manes2021fuzzing}. For statically linked candidates, the linker must also search library archives and extract the required object-file members for each new executable, even though those archives remain unchanged~\cite{gnu-ld-options}. Under either linking mode, building and starting a fresh executable repeats preparation of the same library and runtimes, even though reduction changes only the harness.

The opportunity for reuse is broader than header caching: most of the build and execution environment remains fixed across queries. Exploiting it requires separating that environment from the changing candidate. Execution reuse must also preserve isolation, since a candidate may crash or modify process state in ways that would affect the next query. The challenge is to reuse stable preparation while keeping candidate compilation and execution independent.

\subsection{Input Consumption Dependencies}
\label{sec:challenge-inputs}

Preserving the original crash input does not necessarily preserve the arguments passed to the library. A libFuzzer harness receives an input buffer and its length~\cite{llvm-libfuzzer}, and may either pass these bytes directly to library APIs or consume them through \texttt{FuzzedDataProvider} (FDP)~\cite{llvm-fuzzed-data-provider} to construct numbers, strings, and buffers. Because FDP consumes the input statefully, the value returned by each call depends on how many bytes have already been consumed by preceding calls~\cite{llvm-fuzzed-data-provider}. Removing an earlier FDP operation can therefore shift subsequent consumption and change the arguments supplied to surviving library calls, even when the input itself remains unchanged.

\figref{fig:input-consumption} illustrates this dependency. The first call consumes two bytes whose value is otherwise unused. The second returns the two bytes passed to the library. Deleting the unused call changes the second result from \texttt{\{30, 40\}} to \texttt{\{10, 20\}}. If the original crash requires \texttt{\{30, 40\}}, the interestingness test rejects the deletion. The reducer therefore retains the first call in the harness even though its result is unused. It survives because it advances FDP's consumption state, not because it contributes to the crash.

\begin{figure}[tbp]
  \centering
\begin{lstlisting}[style=paper-cpp]
// Crash input: {10, 20, 30, 40}.
FuzzedDataProvider fdp(data, size);
auto unused  = fdp.ConsumeBytes<uint8_t>(2); // {10, 20}
auto payload = fdp.ConsumeBytes<uint8_t>(2); // {30, 40}
library_api(payload.data(), payload.size());
\end{lstlisting}
  \caption{An illustrative input consumption dependency. Deleting the unused consumption changes \texttt{payload} to \texttt{\{10, 20\}}, even though the crash input is unchanged.}
  \Description{A C++ example with two successive FDP byte consumption calls. The first result is unused. The second is passed to a library API, but deleting the first call changes which two bytes it receives.}
  \label{fig:input-consumption}
\end{figure}

This dependency also arises when consumption sites execute in loops or when calls such as \texttt{remaining\_bytes()} determine control flow. Deleting a consumption operation can then change values in later iterations or alter which library calls execute~\cite{havoc2026}. A useful reduction must decouple the values supplied by surviving sites from the consumption performed by deleted sites. It must also retain enough information to expose those values in the final reproducer. Otherwise, the developer still has to reconstruct the arguments from the original input and its parsing code.

\subsection{Recognizing the Reference Crash Efficiently}
\label{sec:challenge-oracle}

Preserving the original input also does not guarantee that a reduced harness reproduces the same crash. Removing a setup or initialization call may place the library in a different state and trigger an unrelated failure. The \emph{crash oracle}, which serves as the decision rule in the interestingness test, must therefore distinguish candidates that preserve the \emph{reference crash} from those that merely crash for a different reason. Since distinct bugs can manifest as the same error category, existing approaches rely on stack-trace information to capture both the fault location and its calling context. FaST compares function names and their positions when grouping crash reports~\cite{fast2022}, while FuzzerAid matches the bug type, source location, and call stack when classifying crashing inputs~\cite{fuzzeraid2022}.

\begin{figure}[!t]
  \centering
  \textbf{(a) Symbolized crash location and caller}
\begin{lstlisting}[style=paper-trace]
#0 av1_one_pass_cbr_svc_start_layer
   svc_layercontext.c:447:18
#1 av1_get_compressed_data
   encoder.c:5338:5
\end{lstlisting}
  \begin{minipage}[t]{0.49\linewidth}
    \centering
    \textbf{(b) Shared-library offsets}
\begin{lstlisting}[style=paper-trace]
#0 (libaom.so+0x201cf25)
#1 (libaom.so+0x1a9a22e)
\end{lstlisting}
  \end{minipage}\hfill
  \begin{minipage}[t]{0.49\linewidth}
    \centering
    \textbf{(c) Statically linked executable}
\begin{lstlisting}[style=paper-trace]
#0 (harness+0x13a81c5)
#1 (harness+0xe7535e)
\end{lstlisting}
  \end{minipage}
  \caption{Alternative representations of a libaom crash; path prefixes and later frames are omitted. Shared-library offsets refer to a fixed binary, whereas executable offsets can move when a candidate is relinked.}
  \Description{The first two frames of a libaom crash appear as symbolized function names and source locations, as offsets within libaom.so, and as offsets within a statically linked executable.}
  \label{fig:crash-traces}
\end{figure}

Obtaining these function names and source locations adds work to each crashing query. Stack unwinding records instruction addresses, and \emph{symbolization} translates them into function names, files, and line numbers by locating and querying the corresponding binaries and debug information~\cite{llvm-symbolizer}. When LLVM uses an external symbolizer, starting and communicating with a helper process adds to this work~\cite{llvm-asan}. Even a short harness can produce a time-consuming trace through a large library and its dependencies. Reducing the harness does not eliminate this translation for library frames, so many queries repeat this diagnostic work.

Disabling symbolization reduces diagnostic overhead but removes the source-level information used to characterize the crash. In \figref{fig:crash-traces}, function names and file-and-line locations are replaced by module names and numeric offsets, obscuring both the fault location and its calling context. Comparing these numeric entries across candidates is also difficult, as relinking a modified harness can move instructions within the executable, so the same fault may appear at different offsets. If disable symbolization, it bring challenges to distinguish changes caused by harness simplification from those indicating a different bug.

\section{Design}
\label{sec:design}

\begin{figure}[tbp]
  \centering
  \includegraphics[width=\linewidth]{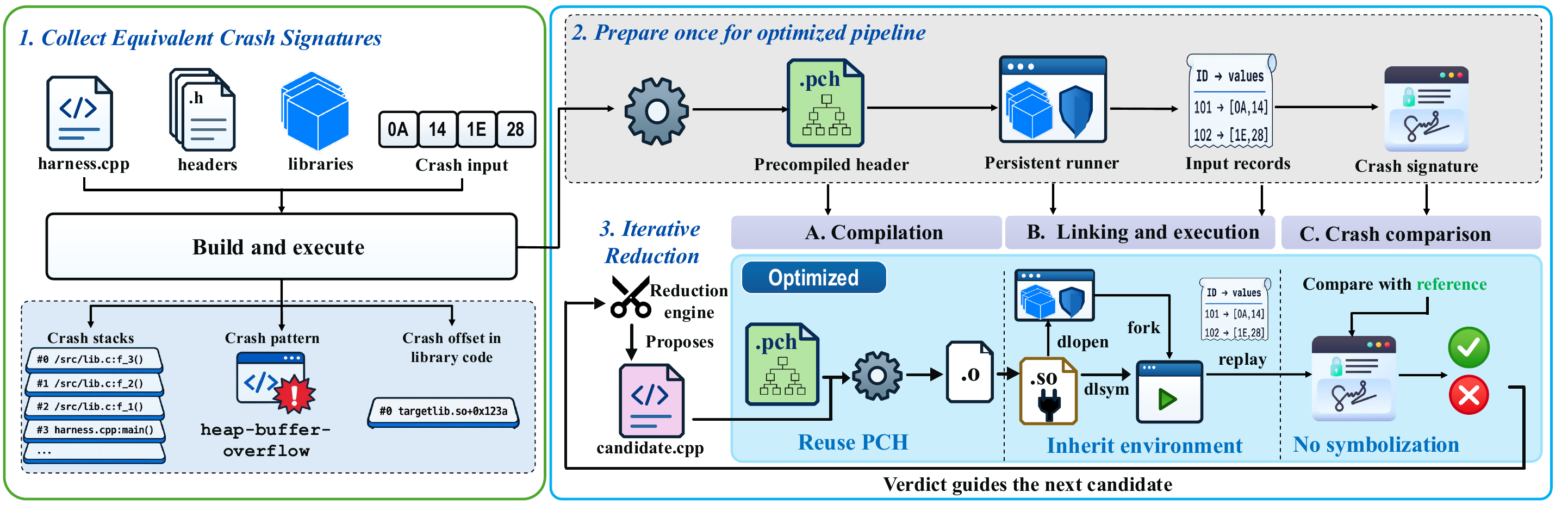}
  \caption{Workflow of \sysname in harness minimization.}
  \Description{The original harness, input, and build options prepare a precompiled header, a runner with libraries and recorded inputs, and crash evidence. A reduction engine sends candidates through compilation, isolated execution with replay, and a crash check, then receives the verdict. Once reduction finishes, inputs are embedded and the reproducer undergoes symbolized validation.}
  \label{fig:design-overview}
\end{figure}

\sysname is designed to efficiently minimize harnesses built on top of existing program reduction engines, while addressing the challenges in \secref{sec:challenges}. The core idea behind \sysname is to reduce the redundant work in harness minimization to enable efficient reduction. It separates the harness code being reduced from the information that should remain stable during reduction: \textit{the build environment}, \textit{the values consumed from the crash input}, and \textit{the reference crash}. \figref{fig:design-overview} shows how these mechanisms cooperate. Given a harness, its crash input, and its header files and libraries used for compilation and linking, \sysname first reproduces the original crash to collect the required crash signatures. Then it pre-compiles the header files and prepares a persistent runner that loads the target libraries that support runtimes to construct a reusable build and execution context, and records the input values consumed by FDP. An existing reduction engine then supplies candidates to a query procedure that compiles the changing code, executes it with the recorded input values, and returns an oracle verdict. When the reduction finishes, \sysname restores the source needed for an ordinary build, embeds the surviving inputs, and validates the resulting reproducer. The reduction engine controls which transformations to try, while \sysname controls how each transformation is evaluated. By integrating these mechanisms, \sysname can significantly improve the efficiency and quality of harness minimization over existing program reduction approaches.

\subsection{Reusing the Build and Execution Environment}
\label{sec:design-environment}

The first mechanism establishes a boundary around the changing harness body. On the compilation side, declarations outside this boundary are prepared once. On the execution side, the library and runtime implementations are prepared once. In the further reduction queries, only the changing harness body incurs additional overhead. A query rebuilds the body and connects it to these prepared components, while a separate child process isolates its execution.

\subsubsection{Header Isolation with Precompilation}
\label{sec:design-pch}

We isolate the stable context in a fixed \emph{header prefix} by separating the headers from the harness code, leaving only a reducible \emph{harness body}.
For each harness, \sysname automatically identifies the prefix by analyzing the include statements from the AST nodes of the harness. The harness body contains the remaining declarations and functions. The prefix is then compiled into a precompiled header (PCH) file, which is then used in subsequent compilations of the harness body and makes only the body available to the reduction engine. Each candidate compilation loads the same PCH, so source transformations do not trigger repeat analysis of the fixed headers. This approach avoids repeatedly parsing and analyzing the same headers for each candidate, significantly reducing compilation time.

This boundary preserves the interface against which candidates are checked. Macro definitions (e.g., \texttt{\#ifdef}) that affect an include remain with that include, and the PCH and candidate builds use compatible compiler settings. The compiler still checks the modified body and performs the instantiation and code generation it requires. Precompiling the isolated prefix therefore removes repeated analysis without bypassing candidate type checking. The generated object code becomes the input to the execution mechanism below.

\subsubsection{Dedicated Execution with the Persistent Runner}
\label{sec:design-linkage}

Precompilation reuses the declarations needed to compile a candidate, but an ordinary build still links the resulting object with the target libraries and starts a new executable for every query. We move the stable library implementations and their initialization outside this loop. A \emph{persistent runner} retains the prepared execution environment, while each candidate becomes a \emph{harness plugin}: a small shared object containing the changing harness code. The plugin leaves references to library APIs unresolved until loading, when they bind to implementations already available in the runner. This separation reuses both the linked library code and the process state established before harness execution.

\begin{figure}[tbp]
  \centering
  \includegraphics[width=0.7\linewidth]{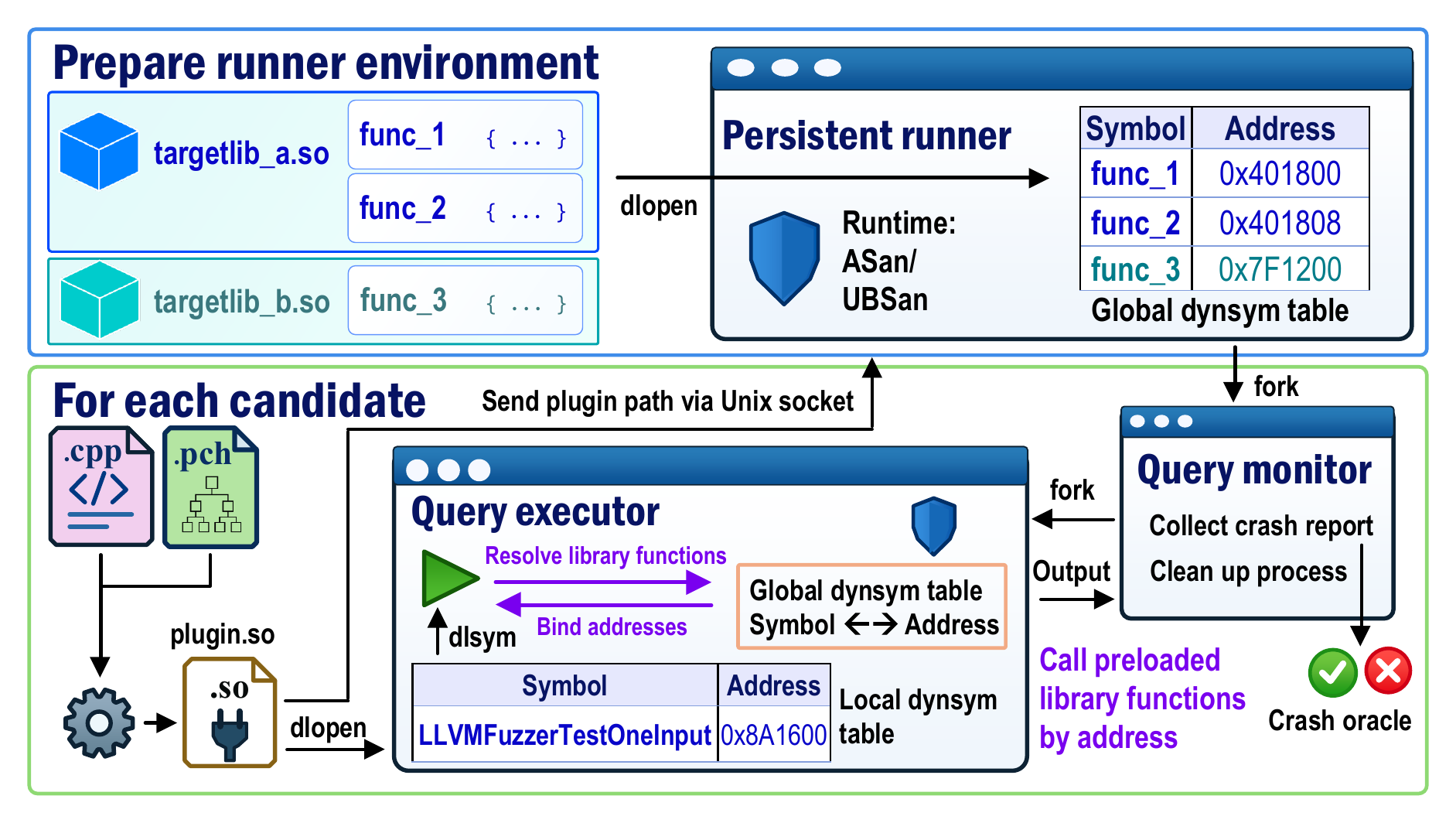}
  \caption{Persistent execution with shared target libraries. The parent prepares the resident environment before serving queries. Each query builds a plugin and creates a monitor and an executor; only the executor loads and runs candidate code. The parent survives crashes and supplies the prepared state to subsequent queries.}
  \Description{The runner loads shared target libraries and prepares sanitizer runtimes. Each candidate is compiled with a precompiled header into a shared library plugin whose path is sent to the runner. The runner forks a query monitor, which forks an executor. The executor loads the plugin, resolves library calls using inherited symbol information, locates the harness entry point, and executes the harness. The monitor collects its output and passes the report to the crash oracle.}
  \label{fig:persistent-runner}
\end{figure}

\textbf{Preparing the resident environment.} \figref{fig:persistent-runner} shows the dynamic-library case. We first identify the target shared libraries retained by the original executable's dependency metadata. Using this set avoids introducing libraries discarded by the original link, whose dependencies or initialization could change execution, such as third-party libraries. The runner retains these dependencies at link time and opens the target libraries with \texttt{dlopen} using \texttt{RTLD\_NOW} and \texttt{RTLD\_GLOBAL}: symbol references are resolved eagerly, and the libraries' exported symbols become available to subsequently loaded plugins~\cite{linux-dlopen}. Library mappings, relocations, and constructors are thus prepared before candidate execution. The runner also initializes supporting runtimes and reads the crash input into memory, then remains alive in a request loop on a Unix-domain socket.

\textbf{Executing a new candidate.} Each query compiles the candidate body as position-independent code, links it into a plugin, and submits the plugin path to the runner. The runner starts a disposable \emph{monitor}, which forks an \emph{executor}. Only the executor loads the plugin, using \texttt{RTLD\_NOW} to bind its external references to the inherited environment and \texttt{RTLD\_LOCAL} to keep its definitions local to that load. It then locates \texttt{LLVMFuzzerTestOneInput} with \texttt{dlsym} and invokes it with the retained input buffer. The monitor collects the executor's output and exit status, enforces a timeout, and cleans up its process group before returning the crash report to the query procedure. 

\textbf{Reusing initialization through process inheritance.} Both forks inherit the prepared address space without replacing it through \texttt{exec}. Linux implements this inheritance with copy-on-write: private memory pages remain physically shared until a process modifies them~\cite{linux-fork}. The executor therefore starts with the library code, loader state, and initialized data already in place, while its writes to private memory do not alter the parent's state. Loading the plugin only after forking also keeps candidate constructors out of the persistent parent. The next query derives from that parent again, so neither a crash nor mutations to library globals carry over through private memory. This fork-server design~\cite{hopper2023} removes repeated executable startup and preparation of the resident libraries. The remaining work in a candidate reduction query leaves only process creation, plugin loading and relocation, and harness execution. Process inheritance reuses library initialization as well as isolates private memory for each execution, so the persistent runner can handle more queries without repeated linking and restarting.

\begin{figure}[tbp]
  \centering
  \begingroup
  \definecolor{staticPrepared}{HTML}{246B50}
  \newcommand{\staticStepNumber}[1]{%
    \tikz[baseline=(step.base)]{%
      \node[circle, fill=staticPrepared, text=white, inner sep=0pt,
        minimum size=3.4mm,
        font=\sffamily\fontsize{7}{8}\selectfont\bfseries]
        (step) {#1};%
    }%
  }
  \begin{tikzpicture}[
    font=\sffamily\fontsize{8}{10}\selectfont,
    box/.style={draw=listingBorder, rounded corners=2pt, align=center,
      inner sep=5pt, line width=0.5pt},
    fixed/.style={box, fill=green!4!white},
    heading/.style={font=\sffamily\fontsize{8}{10}\selectfont\bfseries,
      anchor=west},
    flow/.style={-{Latex[length=1.7mm]}, line width=0.65pt,
      draw=staticPrepared},
    note/.style={font=\sffamily\fontsize{7.5}{9}\selectfont,
      align=center, fill=white, inner sep=2pt}
  ]
    \node[heading, text=staticPrepared] at (0,0) {PREPARE ONCE: ARCHIVES BECOME PART OF THE RUNNER};
    \node[fixed, text width=2.95cm, minimum height=1.1cm]
      (harness) at (1.65,-1.15)
      {\textbf{Original harness}\\Compile analysis object\\Undefined symbols $U$};
    \node[fixed, text width=3.0cm, minimum height=1.1cm]
      (roots) at (6.05,-1.15)
      {\textbf{Required API symbols}\\$R=U\cap D$\\e.g., \texttt{api\_call}};
    \node[fixed, text width=2.95cm, minimum height=1.65cm]
      (archives) at (1.65,-3.35)
      {\textbf{Static archives (\texttt{.a})}\\Defined symbols $D$\\[3pt]
       \texttt{api.o}; \texttt{helper.o}\\
       \textcolor{black!50}{\texttt{unused.o} (unrelated)}};
    \node[fixed, draw=staticPrepared, fill=green!7!white,
      text width=3.85cm, minimum height=1.95cm]
      (link) at (6.05,-3.35)
      {\textbf{Link runner once}\\[4pt]
       \begin{tabular}{@{}c@{\hspace{4pt}}l@{}}
         \staticStepNumber{1} & \textcolor{staticPrepared}{\textbf{Set API visibility}} \\[3pt]
         \staticStepNumber{2} & \textcolor{staticPrepared}{\textbf{Extract $R$ + dependencies}} \\[3pt]
         \staticStepNumber{3} & \textcolor{staticPrepared}{\textbf{Export dynamic symbols}}
       \end{tabular}};
    \draw[flow] (harness) -- (roots);
    \draw[flow] ([yshift=4mm]archives.east) -- (3.85,-2.95)
      -- (3.85,-1.45) -- ([yshift=-3mm]roots.west);
    \draw[flow] (roots) -- node[right, note] {roots} (link);
    \draw[flow] (archives) -- (link);

    \draw[draw=staticPrepared, fill=green!2!white,
      rounded corners=3pt, line width=0.65pt]
      (9.05,-0.6) rectangle (12.95,-4.25);
    \node[font=\sffamily\fontsize{8}{10}\selectfont\bfseries]
      at (11.0,-0.95) {Runner executable};
    \node[box, fill=white, text width=3.05cm, minimum height=1.0cm]
      (code) at (11.0,-1.95)
      {\textbf{Selected library code}\\\texttt{api.o} $\rightarrow$ \texttt{helper.o}\\+ runtime support};
    \node[box, fill=white, text width=3.05cm, minimum height=1.0cm]
      (symbols) at (11.0,-3.35)
      {\textbf{Dynamic symbol table}\\\texttt{api\_call} $\rightarrow$ address};
    \draw[flow] (code) -- (symbols);
    \draw[flow] (link.east) -- (9.05,-3.35);
    \node[align=center, text=black!60, text width=3.25cm]
      at (1.65,-4.7) {Unused members stay out.};
    \node[align=center, text=black!60, text width=3.3cm]
      at (6.05,-4.7) {Use archive copies.};

  \end{tikzpicture}
  \endgroup
  \caption{Preparing the runner from static archives. Symbols referenced by the original harness guide archive extraction; the selected implementation is linked into the runner and exported for dynamic lookup. The analysis object supplies symbol information only. The example shows \texttt{api.o} and its dependency selected while an unrelated member remains excluded.}
  \Description{The original harness is compiled into an analysis object. Its undefined symbols are intersected with definitions in the static archives to obtain extraction roots. The one-time runner link highlights three numbered actions: 1, set API visibility in archive copies; 2, extract the required members and their dependencies; 3, export dynamic symbols. An example selects api.o and helper.o but excludes unused.o.}
  \label{fig:static-library-adaptation}
\end{figure}
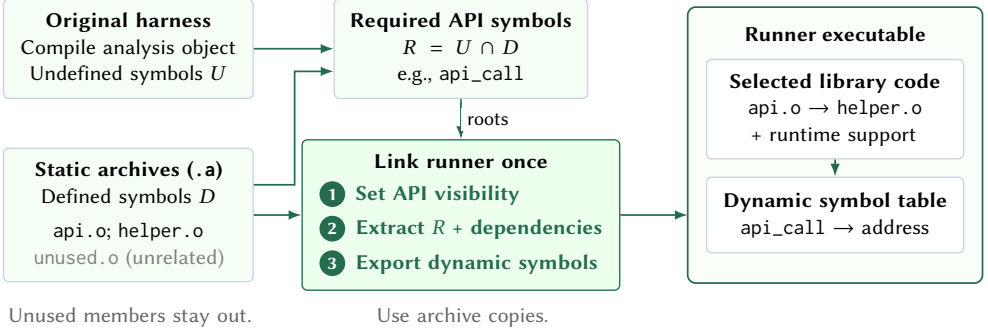

\textbf{Making static libraries reusable.} Direct loading of static libraries in the peristent runner is not feasible because a static archive cannot be opened with \texttt{dlopen}. We mitigate this issue by extracting the archive members and linking the required objects into the runner executable at build time. \figref{fig:static-library-adaptation} shows how we incorporate the required archive members into the runner executable and expose their APIs to plugins. This requires both \emph{selecting the code} and \emph{exporting its symbols}. Merely adding an archive to the runner's link command is insufficient: because the runner does not directly call the harness-specific APIs, the linker may leave their object files unselected. Conversely, including every archive member can introduce unrelated dependencies or constructors that the original harness never required.

To select the required code, we compile the original harness body once into an analysis object. Let $U$ be its undefined symbols and $D$ the symbols defined by the supplied archives. Their intersection $R=U\cap D$ identifies the library symbols referenced by that harness. We use \(R\) as the roots of archive extraction by marking these symbols as unresolved with the linker's \texttt{-u} option. This causes the linker to extract archive members that define the root symbols and transitively include additional members required to resolve newly introduced references. Grouping multiple archives allows such dependencies to be resolved across archive boundaries~\cite{gnu-ld-options}. The analysis object supplies the roots but is not itself linked into the runner, keeping the original harness code and its initialization outside the persistent parent. The resulting runner contains the selected library implementation and is built as a non-PIE executable, allowing reuse of archive objects.

To make the selected library code callable from harness plugins, we expose its API symbols through the runner. As shown in \figref{fig:static-library-adaptation}, we first change required APIs with hidden visibility to default visibility in archive copies (step~\ding{172}), leaving the original libraries unchanged. The runner link then extracts the required archive members and their dependencies (step~\ding{173}) and uses \texttt{--export-dynamic} to expose their externally visible definitions in the dynamic symbol table (step~\ding{174})~\cite{gnu-ld-options}. If extraction roots are unavailable, step~\ding{173} instead includes whole archives. At runtime, plugin API references bind to these exported definitions in the inherited executable. Candidates thus reuse both the library code and its initialized state, avoiding repeated archive extraction, library relocation, and library initialization.

\subsection{Preserving Inputs with Call-Site Replay}
\label{sec:design-replay}

To address the input consumption dependencies described in \secref{sec:challenge-inputs}, \sysname records the values consumed from the crash input and replays them for surviving calls. This decouples the values supplied to the library from the operations that consume them, so a candidate can delete calls that advance FDP's consumption state without changing the values seen by surviving calls.
\begin{figure}[tbp]
  \centering
  \begingroup
  \definecolor{replayKept}{HTML}{246B50}
  \definecolor{replayChanged}{HTML}{A84735}
  \newcommand{\replayDeleted}[1]{%
    \tikz[baseline=(deleted.base)]{%
      \node[inner sep=0pt, outer sep=0pt,
        font=\ttfamily\footnotesize, text=black!55] (deleted) {#1};
      \draw[replayChanged, line width=0.5pt] (deleted.west) -- (deleted.east);
    }%
  }
  \lstdefinestyle{replay-code}{
    style=paper-cpp,
    frame=none,
    backgroundcolor={},
    xleftmargin=0pt,
    xrightmargin=0pt,
    aboveskip=0pt,
    belowskip=0pt,
    escapeinside={(*@}{@*)}
  }
  \newsavebox{\replayOriginalCode}
  \newsavebox{\replayRecordedCode}
  \newsavebox{\replayReducedCode}
  \begin{lrbox}{\replayOriginalCode}
    \begin{minipage}{3.7cm}
\begin{lstlisting}[style=replay-code]
auto unused =
  fdp.ConsumeBytes(2);

auto payload =
  fdp.ConsumeBytes(2);

library_api(payload);
\end{lstlisting}
    \end{minipage}
  \end{lrbox}
  \begin{lrbox}{\replayRecordedCode}
    \begin{minipage}{3.7cm}
\begin{lstlisting}[style=replay-code]
auto unused =
 fdp.ConsumeBytes(2, (*@\textcolor{listingKeyword}{\textbf{101}}@*));

auto payload =
 fdp.ConsumeBytes(2, (*@\textcolor{replayKept}{\textbf{102}}@*));
// payload = {30, 40}
library_api(payload);
\end{lstlisting}
    \end{minipage}
  \end{lrbox}
  \begin{lrbox}{\replayReducedCode}
    \begin{minipage}{3.7cm}
\begin{lstlisting}[style=replay-code]
(*@\replayDeleted{\textbf{auto} unused =}@*)
 (*@\replayDeleted{fdp.ConsumeBytes(2, \textbf{101});}@*)
// redundant call deleted!
auto payload =
 fdp.ConsumeBytes(2, (*@\textcolor{replayKept}{\textbf{102}}@*));

library_api(payload);
\end{lstlisting}
    \end{minipage}
  \end{lrbox}
  \begin{tikzpicture}[
    font=\sffamily\footnotesize,
    heading/.style={font=\sffamily\footnotesize\bfseries, align=center},
    code/.style={draw=listingBorder, rounded corners=2pt,
      inner sep=6pt, minimum height=2.65cm, anchor=north},
    note/.style={align=center, text width=3.7cm},
    result/.style={draw=listingBorder, rounded corners=2pt,
      align=center, text width=3.55cm, minimum height=1.1cm, inner sep=5pt},
    flow/.style={-{Latex[length=1.8mm]}, line width=0.65pt},
    replay/.style={flow, dashed, draw=replayKept}
  ]
    \node[heading] at (0,0) {(a) Original harness};
    \node[heading] at (4.6,0) {(b) Instrument and record};
    \node[heading] at (9.2,0) {(c) Reduce with replay};
    \node[note] at (0,-0.55)
      {Input: \texttt{\{10,20,30,40\}}};
    \node[note] at (4.6,-0.55) {Assign stable call-site IDs};
    \node[note] at (9.2,-0.55) {Preserve IDs after deletion};

    \node[code, fill=listingBackground] (original) at (0,-1)
      {\usebox{\replayOriginalCode}};
    \node[code, fill=blue!3!white] (recorded) at (4.6,-1)
      {\usebox{\replayRecordedCode}};
    \node[code, fill=green!4!white] (reduced) at (9.2,-1)
      {\usebox{\replayReducedCode}};
    \draw[flow] (original.east) -- (recorded.west);
    \draw[flow] (recorded.east) -- (reduced.west);

    \node[note, text=replayChanged] (deletion) at ([yshift=-6mm]original.south)
      {Delete the first call\\without replay};
    \node[result, fill=red!3!white] (changed) at ([yshift=-18mm]original.south)
      {Shared FDP state shifts\\API input becomes \textcolor{replayChanged}{\texttt{\{10,20\}}}};
    \draw[flow, draw=replayChanged] (original.south) -- (deletion.north);
    \draw[flow, draw=replayChanged] (deletion.south) -- (changed.north);

    \node[result, fill=blue!3!white] (trace) at (recorded |- changed) {
      \begin{tabular}{@{}c@{\hspace{10pt}}l@{}}
        \textbf{ID} & \textbf{Recorded result} \\
        \textcolor{listingKeyword}{\texttt{101}} & \texttt{\{10,20\}} \\
        \textcolor{replayKept}{\texttt{102}} & \textcolor{replayKept}{\texttt{\{30,40\}}}
      \end{tabular}
    };
    \draw[flow] (recorded.south) -- node[right, font=\sffamily\scriptsize]
      {execute once} (trace.north);

    \node[result, fill=green!4!white] (preserved) at (reduced |- changed)
      {Replay the record for ID \textcolor{replayKept}{\texttt{102}}\\
       API input stays \textcolor{replayKept}{\texttt{\{30,40\}}}};
    \draw[flow] (reduced.south) -- (preserved.north);
    \draw[replay] (trace.east) -- (preserved.west);
  \end{tikzpicture}
  \endgroup
  \caption{Call-site replay decouples input consumption: after call 101 is deleted, call 102 still supplies \texttt{\{30,40\}} to the API. Code is schematic: \texttt{ConsumeBytes} denotes FDP byte consumption, and added arguments are call-site IDs instrumented by \sysname. The reduction deletes a redundant FDP call.}
  \Description{Three panels show the original harness, the harness instrumented with call-site IDs, and a reduced harness. The original consumes two bytes into an unused value, then two bytes into the payload passed to a library API. Recording associates ID 101 with bytes 10 and 20 and ID 102 with bytes 30 and 40. The reduced harness deletes call 101 and replays the recorded result of call 102. A comparison below shows that deletion without replay changes the API input to bytes 10 and 20, whereas replay preserves bytes 30 and 40.}
  \label{fig:replay-workflow}
\end{figure}
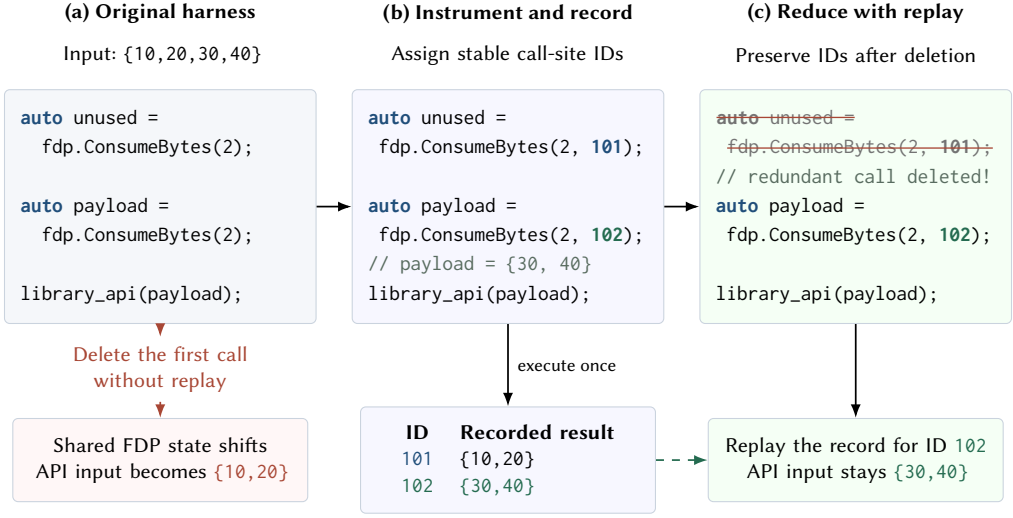

\figref{fig:replay-workflow} illustrates how \sysname binds consumed values to FDP \emph{call sites} to remove input consumption dependencies. In panel~(a), the first call consumes \texttt{\{10,20\}} into an unused variable, while the second supplies \texttt{\{30,40\}} to the library API. Simply deleting the first call shifts the second call's input to \texttt{\{10,20\}} that alters the crash behaviors. To break this dependency, \sysname instruments each call site with a stable ID inserted as an additional argument to record the input bytes consumed by this call, then executes the original harness on the crash input to record the results under their IDs. Panel~(b) associates IDs 101 and 102 with \texttt{\{10,20\}} and \texttt{\{30,40\}}, respectively. During reduction, surviving calls retain their IDs and retrieve the associated records instead of consuming bytes from the shared FDP state. Thus, the candidate in panel~(c) can delete call 101 while call 102 still supplies \texttt{\{30,40\}} to the API. Recording and replaying results by call site allows the reducer to remove operations retained only to advance FDP's consumption state. This association follows the broader idea of preserving relationships between generated values and consuming operations across executions~\cite{havoc2026,greduce2025}.

\textbf{FDP calls inside loops.} A call site inside a loop keeps the same ID across iterations but may return a different value on each execution. \sysname therefore records an ordered sequence $T_s=(v_{s,1},\ldots,v_{s,n_s})$ for each site $s$, rather than a single value. Before executing each candidate, replay resets an independent cursor for every sequence. The $k$th execution of site $s$ returns $v_{s,k}$ and advances only that site's cursor, so deleting another call from the loop does not shift the sequence replayed at $s$. If the candidate executes $s$ fewer times, it uses only a prefix of the recorded sequence; if it requests more values than were recorded, replay terminates the execution instead of consuming new bytes. FDP state queries used in loop conditions, such as \texttt{remaining\_bytes()}, are also assigned IDs and replayed from their own sequences.

\textbf{Reusing the replay runtime.} FDP extensively uses template functions to adapt to different argument types while preserving the same underlying data-consumption logic. To avoid repeatedly compiling FDP's template-based replay logic during reduction, \sysname analyzes the original harness to identify the concrete types used by its FDP calls. We generate explicit instantiations~\cite{cpp23} of the corresponding template functions in a separate compilation unit and compile them, together with trace loading and per-site replay support, into a shared library. The persistent runner links against this library and loads it before forking candidate processes (\secref{sec:design-linkage}), allowing successive queries to reuse the loaded code and fixed initialization while retaining independent replay state in each child. PCH reuse the instantiated FDP header to reduces expensive template resolving and the peristent runner reduces repeated FDP object generation and setup.

\textbf{Input Inlining in the final reproducer.} The final reproducer currently exposes the values consumed by surviving calls via a customized replay runtime, i.e., the instrumented FDP header. To make the reproducer self-contained, as well as reduce customized dependency requirements in bug reports to communities, \sysname inlines the recorded values into the source. Each surviving call is replaced with a literal or container initializer for its recorded value, and any large buffers or repeated sequences with elements larger than 128 are placed in a supporting header.
With the inlined values, the reproducer can be built and executed without the replay runtime. Developers can inspect the concrete input values directly in the source, reducing the additional debugging effort needed to understand the crash.

\subsection{Fast Crash Oracle Without Heavy Symbolization}
\label{sec:design-crash-preservation}
Precise recognition of the reference crash with symbolization is expensive (\secref{sec:challenge-oracle}).
Each query must determine whether the candidate retains the reference crash while keeping diagnostic processing inexpensive. \sysname therefore compares crashes using unsymbolized stack evidence. During reduction, symbolization is disabled while stack unwinding and sanitizer diagnostics remain enabled~\cite{llvm-asan}. The resulting report retains the evidence needed for comparison without repeatedly translating instruction addresses into symbol names and source locations~\cite{llvm-symbolizer}.

\textbf{Collecting crash signatures.} Before reduction, \sysname executes the original harness through the same execution arrangement used for candidates. It records a \emph{crash pattern} from sanitizer warnings about the reported error category or stable diagnostic wording, removing incidental values such as changing addresses. A candidate must crash and match this pattern, then satisfy a stack-based check. For this check, \sysname selects the library offset as oracle for dynamic linking or the stack depth as oracle for static linking bypassing heavy symbolization.

\textbf{Offset oracle for dynamic linking.} We record the instruction offset of the first target-library frame in the reference crash stack reported by the sanitizers. In a candidate report, the first frame from the recorded library must have the same offset. A different offset, or the absence of the required library frame, prevents acceptance even when the crash pattern matches. Because the library binary remains fixed throughout reduction, its internal offsets retain their meaning as the harness changes. The oracle thus compares the observed library location directly, without recovering its function name or source line.

\textbf{Depth oracle for static linking.} When static library code is incorporated into the runner, its frames are reported under the executable rather than a separate target-library module. Therefore, we compare the number of frames in its first crash stack with the corresponding count from the candidate, together with the crash pattern. Both counts are obtained through the same execution arrangement, so runner frames are accounted for consistently. We count only the crash stack, excluding additional allocation or deallocation traces in sanitizer reports. Stack depth provides a coarse structural constraint: different faults can share both an error pattern and a frame count, so this oracle serves as a filtering heuristic during reduction. The collision of different faults with the same depth is possible, but rare. To detect such collision, the final reproducer is validated with symbolization and reports a collision to users. Users can then repeat reduction from the original harness with symbolized crash comparison or switch to dynamic linking when a shared library build is available, as discussed in \secref{sec:discussion}.

\section{Implementation}
\label{sec:implementation}

HarnessMinimizer is implemented primarily in Python and C++, with approximately 9,400 lines of source code. Python coordinates source transformation, compilation, and crash checking, while C++ implements the persistent runner and replay runtime described in \secref{sec:design-linkage} and \secref{sec:design-replay}.

\textbf{Source Transformation and Input Replay.}
We use Tree-sitter's C++ parser~\cite{treesitter} to identify FDP call sites for instrumentation and replace surviving calls with recorded values in the final reproducer. For repeated calls, input inlining emits value sequences and indices; large buffers are placed in a supporting header. For harnesses that consume the input buffer directly, we bind the fuzzing entry point's parameters to the embedded crash bytes. Clang builds libraries with AddressSanitizer~\cite{asan2012} and UndefinedBehaviorSanitizer~\cite{llvm-ubsan} instrumentation.

\textbf{Reduction Engine Integration.}
We integrate treereduce~\cite{treereduce}, Perses~\cite{perses2018}, WDD~\cite{wdd2025}, and CDD~\cite{cdd2025} through a shared interestingness test. Adapters translate its verdict into each engine's acceptance convention, giving all engines the same optimization provided by \sysname.

\section{Evaluation}
\label{sec:evaluation}
To demonstrate the efficiency and effectiveness of \sysname, we evaluate its performance on an extensive set of 200 crash cases across 21 open source libraries. To construct such an evaluation set, we first collected 100 crashes from the artifacts released by PromptFuzz~\cite{promptfuzz2024}, PromeFuzz~\cite{promefuzz2025} and FuzzAgent~\cite{fuzzagent2026}, and confirmed as library bugs by maintainers. Then we complemented them by executing these three systems to collect 100 additional crashes reported as harness bugs and confirmed through manual analysis, which were used as negative samples for crash triage in \secref{sec:evaluation-triage}. Since \sysname does not rely on a specific reduction engine, we evaluate it with four popular program reduction engines: treereduce~\cite{treereduce}, Perses~\cite{perses2018}, WDD~\cite{wdd2025}, and CDD~\cite{cdd2025}. The evaluation examines \sysname on the following aspects: \emph{reduction performance} in \secref{sec:evaluation-performance}, \emph{effectiveness of each component} in \secref{sec:evaluation-components}, and \emph{impact on crash triage} in \secref{sec:evaluation-triage}.

All reduction experiments run in an Ubuntu 24.04.4 LTS environment on a server with two Intel Xeon Platinum 8580 processors (120 physical cores / 240 threads, 2.0\,GHz) and 2~TiB of RAM. We use Clang 21.1.8 and GNU ld 2.42 as the compilation tools.

\subsection{Reduction Performance}
\label{sec:evaluation-performance}

We evaluate the time required to minimize a harness and the resulting source size by integrating \sysname with treereduce~\cite{treereduce}, WDD~\cite{wdd2025}, CDD~\cite{cdd2025}, and Perses~\cite{perses2018}. To ensure consistent evaluation, each engine was configured to use only one worker under a sequential candidate generation and interestingness testing setting, eliminating implementation variance of different parallelization strategies in these systems. In the reduction processes, we didn't impose fixed budget on number of reduction rounds or an overall time. Instead, each engine runs to its own \emph{fixed point} until a complete round finds no further simplification that preserves the reference crash. For each engine, we compare a baseline (Base) and an optimized configuration (Opt) with \sysname on the same 200 cases, using the same original harnesses, crash inputs, and library builds. Base rebuilds and reduces each candidate directly and checks crashes with symbolization enabled. Opt uses all of \sysname's optimization methods (\secref{sec:design}). We measure the total wall-clock time~\cite{ghost-in-the-machine}, including preparation, reduction, input inlining, and final validation, and count the source tokens remaining in the reduced harness. \tabref{tab:overall} reports these results. For readability, it displays the first case from each of the 21 libraries, followed by the minimum, arithmetic mean, median, and maximum over all 200 cases.

\begin{table}[tbp]
  \centering
  \begin{threeparttable}
    \caption{Harness sizes and reduction results for 21 selected cases and the full benchmark.}
    \label{tab:overall}
    \footnotesize
    \definecolor{overallAccent}{gray}{0}
    \definecolor{overallHeader}{gray}{0.90}
    \definecolor{overallOptBg}{gray}{0.95}
    \definecolor{overallOptTimeBg}{gray}{0.88}
    \definecolor{overallOptSizeBg}{HTML}{DFEAF2}
    \definecolor{overallRule}{gray}{0.45}
    \hypersetup{citecolor=black}
    \setlength{\tabcolsep}{1.4pt}
    \renewcommand{\arraystretch}{1.12}
    \sisetup{
      group-digits=false,
      round-mode=places,
      round-precision=0,
      detect-weight=true,
      table-number-alignment=right
    }
    \begin{tabular*}{\linewidth}{@{\extracolsep{\fill}}l
      S[table-format=3.0] S[table-format=4.0]
      S[table-format=5.0] S[table-format=4.0]
      >{\columncolor{overallOptTimeBg}}S[table-format=4.0,number-color=overallAccent]
      >{\columncolor{overallOptSizeBg}}S[table-format=4.0,number-color=overallAccent]
      *{3}{S[table-format=4.0] S[table-format=3.0]
        >{\columncolor{overallOptTimeBg}}S[table-format=4.0,number-color=overallAccent]
        >{\columncolor{overallOptSizeBg}}S[table-format=3.0,number-color=overallAccent]}@{}}
      \arrayrulecolor{overallAccent}
      \toprule
      \rowcolor{overallHeader}
      & \multicolumn{2}{c}{\textcolor{overallAccent}{\textbf{Original}}}
      & \multicolumn{4}{c}{\textcolor{overallAccent}{\textbf{treereduce~\cite{treereduce}}}}
      & \multicolumn{4}{c}{\textcolor{overallAccent}{\textbf{WDD~\cite{wdd2025}}}}
      & \multicolumn{4}{c}{\textcolor{overallAccent}{\textbf{CDD~\cite{cdd2025}}}}
      & \multicolumn{4}{c}{\textcolor{overallAccent}{\textbf{Perses~\cite{perses2018}}}} \\
      \arrayrulecolor{overallRule}
      \cmidrule(lr){2-3} \cmidrule(lr){4-7} \cmidrule(lr){8-11}
      \cmidrule(lr){12-15} \cmidrule(l){16-19}
      & & & \multicolumn{2}{c}{Base} & \multicolumn{2}{>{\columncolor{overallOptBg}}c}{\textcolor{overallAccent}{\textbf{Opt}}}
      & \multicolumn{2}{c}{Base} & \multicolumn{2}{>{\columncolor{overallOptBg}}c}{\textcolor{overallAccent}{\textbf{Opt}}}
      & \multicolumn{2}{c}{Base} & \multicolumn{2}{>{\columncolor{overallOptBg}}c}{\textcolor{overallAccent}{\textbf{Opt}}}
      & \multicolumn{2}{c}{Base} & \multicolumn{2}{>{\columncolor{overallOptBg}}c}{\textcolor{overallAccent}{\textbf{Opt}}} \\
      \cmidrule(lr){4-5} \cmidrule(lr){6-7} \cmidrule(lr){8-9} \cmidrule(lr){10-11}
      \cmidrule(lr){12-13} \cmidrule(lr){14-15} \cmidrule(lr){16-17} \cmidrule(l){18-19}
      \textbf{Crash case} & {LoC} & {Tokens}
      & {Time} & {Size} & {\textcolor{overallAccent}{Time}} & {\textcolor{overallAccent}{Size}}
      & {Time} & {Size} & {\textcolor{overallAccent}{Time}} & {\textcolor{overallAccent}{Size}}
      & {Time} & {Size} & {\textcolor{overallAccent}{Time}} & {\textcolor{overallAccent}{Size}}
      & {Time} & {Size} & {\textcolor{overallAccent}{Time}} & {\textcolor{overallAccent}{Size}} \\
      \midrule
      c-ares-1 & 258 & 1342 & 1181 & 403 & 218 & 392 & 1307 & 251 & 280 & 240 & 1062 & 251 & 232 & 240 & 1107 & 253 & 241 & 242 \\
      cjson-1 & 280 & 1263 & 10276 & 343 & 1812 & 324 & 706 & 159 & 159 & 148 & 611 & 159 & 142 & 148 & 581 & 161 & 137 & 150 \\
      curl-2 & 187 & 848 & 475 & 69 & 102 & 52 & 234 & 35 & 73 & 31 & 219 & 35 & 71 & 31 & 182 & 35 & 63 & 31 \\
      exiv2-1 & 37 & 192 & 269 & 173 & 106 & 172 & 247 & 78 & 92 & 55 & 219 & 78 & 84 & 55 & 223 & 78 & 85 & 55 \\
      lcms-1 & 329 & 1619 & 2185 & 236 & 385 & 236 & 585 & 66 & 207 & 59 & 546 & 66 & 207 & 59 & 519 & 66 & 177 & 59 \\
      libaom-1 & 263 & 1574 & 1345 & 309 & 289 & 279 & 1405 & 242 & 368 & 231 & 1187 & 242 & 324 & 231 & 1156 & 242 & 307 & 231 \\
      libjpeg-turbo-1 & 182 & 1106 & 694 & 132 & 148 & 124 & 458 & 56 & 123 & 28 & 412 & 56 & 114 & 28 & 402 & 56 & 110 & 28 \\
      liblouis-1 & 276 & 1195 & 1432 & 344 & 307 & 340 & 943 & 218 & 234 & 201 & 828 & 218 & 206 & 201 & 832 & 230 & 203 & 213 \\
      libmagic-1 & 487 & 2613 & 3598 & 289 & 699 & 259 & 643 & 92 & 151 & 81 & 583 & 92 & 141 & 81 & 550 & 92 & 132 & 81 \\
      libpcap-1 & 179 & 948 & 9621 & 382 & 1817 & 327 & 532 & 112 & 130 & 95 & 515 & 112 & 126 & 95 & 528 & 124 & 129 & 107 \\
      libpng-1 & 273 & 1090 & 11250 & 474 & 2086 & 444 & 1525 & 677 & 338 & 667 & 1394 & 677 & 309 & 667 & 1389 & 683 & 278 & 673 \\
      libtiff-1 & 162 & 692 & 1116 & 277 & 213 & 241 & 700 & 154 & 142 & 132 & 590 & 154 & 127 & 132 & 629 & 154 & 119 & 143 \\
      libvpx-1 & 161 & 992 & 1061 & 243 & 203 & 194 & 552 & 114 & 135 & 100 & 582 & 114 & 141 & 100 & 388 & 88 & 93 & 77 \\
      opencv-1 & 295 & 1874 & 6764 & 535 & 1867 & 508 & 2356 & 349 & 1188 & 334 & 2206 & 349 & 1136 & 334 & 3045 & 351 & 1078 & 334 \\
      openssl-1 & 319 & 1491 & 2185 & 195 & 271 & 137 & 421 & 77 & 87 & 60 & 409 & 77 & 90 & 60 & 379 & 77 & 83 & 60 \\
      protobuf-1 & 333 & 1489 & 1182 & 158 & 269 & 129 & 780 & 70 & 207 & 39 & 711 & 70 & 194 & 39 & 692 & 70 & 188 & 39 \\
      pugixml-1 & 213 & 1287 & 851 & 168 & 182 & 163 & 679 & 312 & 173 & 307 & 607 & 312 & 159 & 307 & 577 & 312 & 150 & 307 \\
      re2-1 & 349 & 1427 & 1152 & 284 & 230 & 274 & 817 & 125 & 181 & 114 & 748 & 125 & 174 & 114 & 672 & 115 & 154 & 104 \\
      sqlite-1 & 447 & 2015 & 1490 & 571 & 307 & 539 & 1483 & 189 & 370 & 174 & 1283 & 189 & 312 & 174 & 1040 & 214 & 314 & 203 \\
      tinygltf-1 & 351 & 1542 & 2461 & 288 & 1067 & 274 & 1368 & 146 & 720 & 146 & 1057 & 115 & 376 & 101 & 932 & 116 & 679 & 110 \\
      zlib-1 & 194 & 896 & 1035 & 346 & 211 & 324 & 1086 & 136 & 213 & 115 & 1029 & 136 & 183 & 115 & 871 & 136 & 172 & 115 \\
      \midrule
      \rowcolor{overallHeader}
      \multicolumn{19}{l}{\textcolor{overallAccent}{\textbf{Summary over all 200 cases}}} \\
      Min & 35 & 138 & 50 & 23 & 39 & 23 & 49 & 21 & 44 & 20 & 47 & 21 & 42 & 20 & 46 & 21 & 41 & 20 \\
      \bfseries Mean & \bfseries 273.275 & \bfseries 1351.715 & \bfseries 2407.884 & \bfseries 283.205 & \cellcolor{overallOptTimeBg}\bfseries 533.565 & \cellcolor{overallOptSizeBg}\bfseries 263.095 & \bfseries 974.452 & \bfseries 156.495 & \cellcolor{overallOptTimeBg}\bfseries 275.847 & \cellcolor{overallOptSizeBg}\bfseries 143.21 & \bfseries 871.08 & \bfseries 156.615 & \cellcolor{overallOptTimeBg}\bfseries 248.831 & \cellcolor{overallOptSizeBg}\bfseries 143.175 & \bfseries 807.899 & \bfseries 156.02 & \cellcolor{overallOptTimeBg}\bfseries 233.436 & \cellcolor{overallOptSizeBg}\bfseries 142.645 \\
      Median & 270 & 1272 & 1321.648 & 269.5 & 288.135 & 252 & 821.8 & 136 & 202.668 & 118 & 719.95 & 135.5 & 189.628 & 116.5 & 671.388 & 135.5 & 172.453 & 117 \\
      Max & 839 & 4740 & 33989 & 1084 & 8739 & 1063 & 3981 & 745 & 2511 & 731 & 5935 & 745 & 2010 & 731 & 3429 & 745 & 2002 & 731 \\
      \arrayrulecolor{overallAccent}
      \bottomrule
    \end{tabular*}
    \begin{tablenotes}[flushleft]
      \footnotesize
      \item \textit{Notes.} \textbf{LoC}: line of code; \textbf{Tokens} / \textbf{Size}: original / remaining source tokens; \textbf{Time}: total wall time (s).
      \textbf{Base} / \textbf{Opt}: baseline / optimization with \sysname. Values are rounded to integers.
      Opt shading: gray for Time, blue-gray for Size.
    \end{tablenotes}
  \end{threeparttable}
\end{table}

\textbf{Reduction time.} \sysname reduces mean total wall time for all four engines. For treereduce, the mean falls from 2,408 to 534 seconds; for WDD, from 974 to 276 seconds; for CDD, from 871 to 249 seconds; and for Perses, from 808 to 233 seconds. Dividing the baseline mean by the optimized mean \emph{gives speedups of 4.51$\times$, 3.53$\times$, 3.50$\times$, and 3.46$\times$}, respectively. Median times also decrease for every engine, from 671--1,322 seconds under four base engines to 172--288 seconds under optimization with \sysname. The maximum observed time for treereduce decreases from 33,989 to 8,739 seconds, shortening the longest run from approximately 9.4 to 2.4 hours. Minimum times also decrease for every engine, from 46--50 seconds under Base to 39--44 seconds under \sysname. These results demonstrate that \sysname can consistently reduce the time required to minimize a harness across different engines and achieve improvement over all cases, with the largest speedup of 4.51$\times$ observed in treereduce~\cite{treereduce}.

\textbf{Reduction size.} The original harnesses contain 1,352 source tokens on average. With \sysname, the mean remaining size is 263 tokens for treereduce and approximately 143 tokens for each of WDD, CDD, and Perses, achieving a mean reduction rate of 80.55\% for treereduce and 89.43\% for the other engines. Compared with their baseline settings, \sysname 
decreases the mean remaining size range from 7.1\% to 8.9\%. For the largest case with the original size of 4,740 tokens, \sysname reduced 77.57\% tokens on treereduce and 84.57\% on the other engines. For the smallest case with the original size of 138 tokens, \sysname still reduced 83.33\% tokens on treereduce and 85.51\% on the other engines. 
These results show that \sysname can consistently reduce the size of minimized harnesses across different engines and achieve improvement over all cases, with the largest reduction rate of 89.43\% observed in Perses~\cite{perses2018}.

\subsection{Effectiveness of Each Component}
\label{sec:evaluation-components}
\label{sec:evaluation-replay}
\label{sec:evaluation-oracle}

We evaluate how \sysname's components affect candidate testing cost, harness size, and crash preservation. For the timing study, we use the same 200 cases, four engines, Base/Opt configurations, and sequential execution settings as in \secref{sec:evaluation-performance}, running each engine to its fixed point. We measure compilation, linking, and execution times separately. Compilation averages include candidates that build successfully. Linking and execution averages include queries that reach the respective stage. Execution time includes diagnostic reporting. Throughput counts queries per second over the reduction search. For each metric, we first compute the average within each case and then average over all 200 cases. The speedups in \tabref{tab:ablation} are ratios of these means, using Base/Opt for times and Opt/Base for throughput. This breakdown measures the combined effect of the optimizations active at each stage. We then use treereduce to evaluate input replay and crash-oracle decisions separately.

\begin{table}[!htbp]
  \centering
  \begin{threeparttable}
    \caption{Candidate testing costs and throughput across four reduction engines.}
    \label{tab:ablation}
    \small
    \definecolor{ablationHeader}{gray}{0.90}
    \definecolor{ablationOpt}{HTML}{DFEAF2}
    \definecolor{ablationRule}{gray}{0.55}
    \setlength{\tabcolsep}{2pt}
    \renewcommand{\arraystretch}{1.12}
    \sisetup{
      group-digits=false,
      round-mode=places,
      round-precision=2,
      round-pad=true,
      detect-weight=true,
      table-column-width={\dimexpr(0.76\linewidth-16\tabcolsep)/6\relax}
    }
    \begin{tabular}{p{0.13\linewidth}p{0.11\linewidth}
      S[table-format={4.2{$\times$}}]
      *{3}{S[table-format={3.2{$\times$}}]}
      *{2}{S[table-format={1.2{$\times$}}]}}
      \toprule
      \rowcolor{ablationHeader}
      & & \multicolumn{4}{c}{\textbf{Stage time (ms)}}
      & \multicolumn{2}{c}{\textbf{Throughput (queries/s)}} \\
      \arrayrulecolor{ablationRule}
      \cmidrule(lr){3-6} \cmidrule(l){7-8}
      \textbf{Engine} & \textbf{Statistic}
      & {Compile} & {\shortstack{Dynamic\\link}} & {\shortstack{Static\\link}} & {Execute}
      & {\shortstack{Dynamic\\link}}
      & {\shortstack{Static\\link}} \\
      \midrule
      treereduce & Base & 1065.23 & 195.62 & 742.67 & 175.68 & 1.31 & 0.70 \\
      \rowcolor{ablationOpt}
       & \textbf{Opt} & 111.52 & 67.16 & 64.27 & 50.37 & 4.75 & 3.60 \\
       & \textbf{Speedup} & \bfseries 9.55{$\times$} & \bfseries 2.91{$\times$} & \bfseries 11.56{$\times$} & \bfseries 3.49{$\times$} & \bfseries 3.63{$\times$} & \bfseries 5.14{$\times$} \\
      \midrule
      Perses & Base & 1020.45 & 187.35 & 694.61 & 261.32 & 1.53 & 0.73 \\
      \rowcolor{ablationOpt}
       & \textbf{Opt} & 105.43 & 66.29 & 63.94 & 108.93 & 4.63 & 3.92 \\
       & \textbf{Speedup} & \bfseries 9.68{$\times$} & \bfseries 2.83{$\times$} & \bfseries 10.86{$\times$} & \bfseries 2.40{$\times$} & \bfseries 3.03{$\times$} & \bfseries 5.37{$\times$} \\
      \midrule
      WDD & Base & 1010.38 & 187.04 & 687.16 & 261.28 & 1.56 & 0.75 \\
      \rowcolor{ablationOpt}
       & \textbf{Opt} & 105.5 & 66.09 & 63.67 & 112.26 & 4.7 & 3.96 \\
       & \textbf{Speedup} & \bfseries 9.58{$\times$} & \bfseries 2.83{$\times$} & \bfseries 10.79{$\times$} & \bfseries 2.33{$\times$} & \bfseries 3.01{$\times$} & \bfseries 5.28{$\times$} \\
      \midrule
      CDD & Base & 1025.04 & 190.64 & 699.47 & 289.55 & 1.49 & 0.73 \\
      \rowcolor{ablationOpt}
       & \textbf{Opt} & 108.73 & 67.47 & 64.19 & 115.49 & 4.49 & 3.91 \\
       & \textbf{Speedup} & \bfseries 9.43{$\times$} & \bfseries 2.83{$\times$} & \bfseries 10.90{$\times$} & \bfseries 2.51{$\times$} & \bfseries 3.01{$\times$} & \bfseries 5.36{$\times$} \\
      \arrayrulecolor{black}
      \bottomrule
    \end{tabular}
    \begin{tablenotes}[flushleft]
      \footnotesize
      \item \textit{Notes.} Base and Opt use the configurations in \secref{sec:evaluation-performance} and report arithmetic averages over 200 cases. Speedups use the reported averages: Base/Opt for times and Opt/Base for throughput.
    \end{tablenotes}
  \end{threeparttable}
\end{table}

\textbf{Overhead Breakdown.} To evaluate how each component of \sysname contributes to the overall performance improvement, we executed one candidate testing cycle for each of the four engines, and measured the time spent in compilation, linking, and execution stages. \tabref{tab:ablation} shows the detailed overhead breakdown and the corresponding speedups. \emph{Compilation achieves speedups of 9.43--9.68$\times$}. For example, the mean compilation time for treereduce decreases from 1,065.23 to 111.52\,ms, showing that \sysname can reduce significant cost of repeated header analysis through PCH, which leaves less code to process for each candidate (\secref{sec:design-pch}). \emph{Linking achieves speedups of 2.83--2.91$\times$ under dynamic linking and 10.79--11.56$\times$ under static linking.} The persistent runner retains the target libraries and supporting runtimes, so successive candidates require only a small harness plugin to be linked (\secref{sec:design-linkage}). \emph{Mean execution time decreases by factors of 2.33--3.49$\times$}. Forking the prepared runner and using the unsymbolized crash oracle reduce repeated process initialization and diagnostic processing (\secref{sec:design-crash-preservation}). Together, these optimizations yield query throughput speedups of 3.01--3.63$\times$ under dynamic linking and 5.14--5.37$\times$ under static linking across the four engines.

\textbf{Input recording and replay.} We evaluate whether preserving consumed values enables more extensive reduction. We take treereduce as a research target and compare the reduction quality with and without replay (\secref{sec:design-replay}). Of the 200 cases, 166 use FDP for dynamic bytes consumption. We therefore measure remaining source tokens and explicit FDP calls before input inlining procedure when reaching the fixed point. Each call site is counted once, including sites inside loops. \tabref{tab:evaluation-replay} shows that replay decreases mean remaining size from 391.48 to 327.49 tokens, a 16.35\% reduction. The mean percentage of original tokens removed increases from 71.66\% to 75.96\%, or 4.30 percentage points. The effect on input operations is larger: mean remaining FDP calls decrease from 9.10 to 5.24, a 42.38\% reduction. These results support recording consumed values can remove needless FDP calls used only to preserve the consumption state, thus achieving better reduction quality.

\begin{table}[!htbp]
  \centering
  \begin{threeparttable}
    \caption{Effect of input replay on harness size before input inlining.}
    \label{tab:evaluation-replay}
    \small
    \definecolor{replayHeader}{gray}{0.90}
    \definecolor{replayOpt}{HTML}{DFEAF2}
    \definecolor{replayRule}{gray}{0.55}
    \setlength{\tabcolsep}{4pt}
    \renewcommand{\arraystretch}{1.15}
    \sisetup{
      group-digits=false,
      round-mode=places,
      round-precision=2,
      round-pad=true,
      detect-weight=true,
      table-column-width={\dimexpr(0.76\linewidth-12\tabcolsep)/5\relax}
    }
    \begin{tabular}{p{0.24\linewidth}
      S[table-format=4.2]
      S[table-format=3.2]
      *{3}{S[table-format=2.2]}}
      \toprule
      \rowcolor{replayHeader}
      & \multicolumn{3}{c}{\textbf{Source tokens}}
      & \multicolumn{2}{c}{\textbf{FDP calls}} \\
      \arrayrulecolor{replayRule}
      \cmidrule(lr){2-4} \cmidrule(l){5-6}
      \textbf{Configuration}
      & {Original} & {Remaining $\downarrow$}
      & {\shortstack{Removed\\(\%) $\uparrow$}}
      & {Original} & {Remaining $\downarrow$} \\
      \midrule
      Without replay & 1497.02 & 391.48 & 71.66 & 28.07 & 9.10 \\
      \rowcolor{replayOpt}
      \textbf{With replay} & 1497.02 & \bfseries 327.49 & \bfseries 75.96 & 28.07 & \bfseries 5.24 \\
      \arrayrulecolor{black}
      \bottomrule
    \end{tabular}
    \begin{tablenotes}[flushleft]
      \footnotesize
      \item \textit{Notes.} Values are arithmetic means over 166 FDP cases.
      $\downarrow$/$\uparrow$ indicate that lower/higher values are better, respectively.
    \end{tablenotes}
  \end{threeparttable}
\end{table}

\textbf{Crash oracle.} We evaluate whether \sysname's oracles preserve the original crash while avoiding repeated symbolization during candidate testing. Under dynamic linking, the \emph{offset oracle} matches the crash pattern and the instruction offset within the target library. When only static libraries are available, the \emph{depth oracle} matches the pattern and the number of frames in the unsymbolized stack trace (\secref{sec:design-crash-preservation}). Reducing all 200 cases with optimized treereduce produces 551,812 queries. For each query, we record the judgments of the offset, depth, and symbolized oracles, then compare each unsymbolized oracle's verdict with the symbolized verdict. \tabref{tab:evaluation-oracle} reports disagreements and the number of cases in which manual review of the final reproducers identified a changed crash.

\begin{table}[!htbp]
  \centering
  \begin{minipage}[t]{0.53\linewidth}
    \vspace{0pt}
  \begin{threeparttable}
    \caption{Crash oracle comparison.}
    \label{tab:evaluation-oracle}
    \small
    \definecolor{oracleHeader}{gray}{0.90}
    \setlength{\tabcolsep}{2pt}
    \renewcommand{\arraystretch}{1.05}
    \begin{tabularx}{\linewidth}{@{}l*{4}{>{\raggedleft\arraybackslash}X}@{}}
      \toprule
      \rowcolor{oracleHeader}
      \textbf{Oracle}
      & \shortstack{\textbf{Total}\\\textbf{queries}}
      & \shortstack{\textbf{Mismatch}\\\textbf{queries}}
      & \shortstack{\textbf{Mismatch}\\\textbf{cases}}
      & \shortstack{\textbf{Changed}\\\textbf{cases}}\\
      \midrule
      Offset
      & 551,812 & 15,243 & 14 & 0 \\
      Depth
      & 551,812 & 16,872 & 21 & 4 \\
      \bottomrule
    \end{tabularx}
    \begin{tablenotes}[flushleft]
      \footnotesize
      \item Offset and depth apply to dynamic and static linking, respectively. Mismatches are disagreements with symbolized comparison. The last column counts cases whose final reproducers trigger a different crash, as confirmed by manual review.
    \end{tablenotes}
  \end{threeparttable}

  \end{minipage}\hfill
  \begin{minipage}[t]{0.45\linewidth}
    \vspace{0pt}
  \begin{threeparttable}
    \caption{Crash triage accuracy.}
    \label{tab:evaluation-triage}
    \small
    \definecolor{triageHeader}{gray}{0.90}
    \definecolor{triageReduced}{HTML}{DFEAF2}
    \setlength{\tabcolsep}{2pt}
    \renewcommand{\arraystretch}{1.05}
    \begin{tabularx}{\linewidth}{@{}l*{3}{>{\raggedleft\arraybackslash}X}@{}}
      \toprule
      \rowcolor{triageHeader}
      \textbf{Harness}
      & \shortstack{\textbf{Mean}\\\textbf{tokens}}
      & \shortstack{\textbf{Library}\\\textbf{bugs}}
      & \shortstack{\textbf{Harness}\\\textbf{bugs}} \\
      \midrule
      Original & 1352 & 100\% & 28\% \\
      treereduce & 263 & 100\% & \cellcolor{triageReduced}\textbf{44\%} \\
      Perses & 143 & 100\% & \cellcolor{triageReduced}\textbf{51\%} \\
      \bottomrule
    \end{tabularx}
    \begin{tablenotes}[flushleft]
      \footnotesize
      \item Mean source tokens; accuracy over 100 cases per bug category; for treereduce and Perses, we use harnesses reduced by \sysname.
    \end{tablenotes}
  \end{threeparttable}

  \end{minipage}
\end{table}

The offset and symbolized oracles return the same verdict for \emph{97.24\% of queries}. They disagree on 15,243 queries across 14 cases. In 6,544 of these queries, covering four cases, the candidate trace contains the same function call sequence as the reference, but source line numbers in caller frames differ. In the other 8,699 queries across ten cases, the function call sequence also differs. Manual review confirms that the final reproducers in all 14 cases retain the original crash. These results support using the library crash location to guide reduction even when the calling context changes, allowing \sysname to avoid repeated symbolization during candidate testing.

The depth and symbolized oracles return the same verdict for \emph{96.94\% of queries}. They disagree on 16,872 queries across 21 cases. Of these disagreements, 15,287 queries arise from 17 cases with differences in source locations or function call sequences, but manual inspection finds that the final reproducers still trigger the original crash. The remaining 1,585 disagreements arise from four cases whose final reproducers trigger crashes at different locations while still matching the reference pattern and stack depth. The high agreement supports using stack depth for crash comparison under static linking, while the offset oracle provides more precise location evidence for preserving the original crash when stable offsets are available.

\subsection{Impact on Crash Triage}
\label{sec:evaluation-triage}

We evaluate whether reduced harnesses help an LLM distinguish library defects from harness errors. We compare three configurations on the same 200 crashes (including 100 library bugs and 100 harness bugs) used in \secref{sec:evaluation-performance}: the original harness without reduction, the harness reduced by \sysname with treereduce, and the harness reduced by \sysname with Perses. We follow a one-shot LLM-based crash validation to triage these crashes by providing the harness code and its corresponding crash stacks to LLMs, which is used in PromeFuzz~\cite{promefuzz2025}. We use GLM-5.3~\cite{glm5team2026glm5vibecodingagentic} with high reasoning effort as the judge model. We evaluate each crash case independently on 20 times to reduce the effect of randomness in LLM inference. The final classification is determined by the majority vote of the 20 results and compared with the ground truth. For 10 vs. 10 voting cases, we verify the classification manually.

\tabref{tab:evaluation-triage} reports the mean harness size in source tokens and classification accuracy separately for library bugs and harness bugs. Each accuracy is the fraction of the 100 cases in that category classified correctly. Library-bug accuracy remains at 100\% across all three configurations. In contrast, \emph{harness-bug accuracy increases from 28\% with the original harnesses to 44\% with treereduce and 51\% with Perses}, gains of 16 and 23 percentage points, respectively. Thus, minimization of harness code effectively improves the accuracy for triaging library fuzzing crashes. Especially for crashes caused by harness bugs, the triage on the reduced harnesses with Perses achieve 82.14\% more accurate classification than the original harnesses, which can help developers quickly identify the root cause of the crash and find the valid ones among the library crash candidates.

These results are consistent with the LLM favoring a plausible library-bug explanation when a crash manifests inside library code. That explanation fits actual library bugs, but API misuse in the harness can produce similar symptoms. In an unreduced harness, extensive initialization and unrelated API operations may obscure the relevant misuse. Reduction concentrates the supplied context on the calls and inlined input values that reproduce the crash, making the relevant API usage easier to inspect. This offers a plausible explanation for the improved identification of harness bugs while library-bug accuracy remains unchanged.

\section{Discussion}
\label{sec:discussion}

\textbf{Implications.} The improvements across four reduction engines show that reducing repeated work in candidate testing can benefit different reduction strategies (\secref{sec:evaluation-performance}). Separating the changing harness from its reusable build and execution environment allows these improvements without requiring a new reduction algorithm. Input replay further supports simplification by allowing the removal of operations retained only to advance input consumption. The resulting reproducers expose the remaining API calls and their concrete arguments, making the relevant library usage easier to inspect. The improved identification of harness bugs in our LLM study (\secref{sec:evaluation-triage}) provides evidence that this simplification can also assist crash triage, supporting the use of harness minimization between crash discovery and investigation.

\textbf{Limitation.} The usefulness of these reproducers depends on retaining the reference crash. Under static linking, the depth oracle compares the crash pattern and frame count, providing less location information than a library offset or a symbolized trace. Manual review identified collisions in four of the 200 final reproducers evaluated with this oracle (2\%). Three crashed in different library functions, and one crashed in replay code, while the reference pattern and stack depth remained unchanged (\tabref{tab:evaluation-oracle}). These cases illustrate the tradeoff between inexpensive candidate testing and precise crash discrimination. The final symbolized validation described in \secref{sec:design-crash-preservation} checks for such changes after reduction. If it detects a collision, users can repeat reduction from the original harness with symbolized crash comparison. When a shared library build is available, the offset oracle provides more precise location evidence while retaining the benefit of avoiding repeated symbolization.

\textbf{Threats to validity.} Timing measurements can vary with CPU scheduling and contention from other tasks, while LLM classifications can vary across repeated requests. We improve comparability by using the same cases and build environment with one worker per reduction run. For triage, we reduce the influence of individual responses through 20 classifications and majority voting, with tied votes resolved manually. The magnitude of the observed improvements may also vary across settings. Different hardware and toolchains change the relative costs of candidate testing, while different models, prompts, and inference settings may change the benefit of minimization for crash triage.

\section{Related Work}
\label{sec:related-work}

\subsection{Delta Debugging and Program Reduction}
\label{sec:related-reduction}

Delta debugging established a general approach to simplifying test cases through queries that test a designated property~\cite{zeller2002}. Subsequent work improves reduction through richer representations and transformations. HDD applies reduction to hierarchical input structures~\cite{hdd2006}, while Perses uses a language grammar to retain syntactic validity during simplification~\cite{perses2018}. C-Reduce combines transformations specialized for C/C++, including simplification of declarations and expressions~\cite{creduce2012}. Vulcan alternates a main reducer with transformations derived from the language grammar, including identifier and subtree replacement~\cite{vulcan2023}. These transformations can preserve the tested property without immediately reducing program size, enabling the main reducer to continue beyond a local minimum. SFC converts between grammatical forms to expose additional simplifications~\cite{sfc2025}, and Latra provides matching and rewriting templates for expressing transformations tailored to a language~\cite{latra2025}. These approaches expand the changes an engine can consider while leaving acceptance to its property checker.

The cost of queries has also motivated improvements in candidate selection and reuse of test outcomes. ProbDD estimates which elements need to remain and updates its estimates from previous test results to guide subsequent deletions~\cite{probdd2021}. WDD assigns each removable fragment a weight equal to its token count, giving a large function more weight than a short declaration~\cite{wdd2025}. It uses these weights to form partitions containing similar amounts of code or to select deletions that maximize the expected number of tokens removed. CDD simplifies the scheduling behavior underlying ProbDD after identifying the importance of avoiding unproductive queries~\cite{cdd2025}. RCC stores tested candidates in a compact representation so their results can be reused~\cite{reductioncaching2024}. It also discards entries for candidates that cannot recur after further deletions, reducing the memory needed for caching.

HarnessMinimizer integrates existing reduction engines with an interestingness test designed for library fuzzing harnesses. The engines propose source transformations, while HarnessMinimizer tests the resulting candidates against the reference crash and preserves recorded input values. Its optimizations lower the cost of queries and complement improvements in transformations, candidate selection, and caching.

\subsection{Crash Triage and Crash Preservation}
\label{sec:related-crashes}

Crash comparison is needed both to group reports of the same bug and to decide whether a reduction candidate preserves the crash under investigation. FaST compares stack traces by matching functions and using their positions and frequency across reports to score similarity~\cite{fast2022}. Its comparison time grows in proportion to the total number of frames in the two traces. FuzzerAid uses program reduction to obtain executable evidence for crash grouping~\cite{fuzzeraid2022}. It collects the execution trace of a crashing input, constructs an executable program from the corresponding source code, and applies C-Reduce to obtain a small \emph{fault signature} that retains the crash location and symptoms. Other crashing inputs are executed on this signature and grouped when they reproduce the same bug type, source location, and call stack. Its reduction process therefore depends on checking that the simplified program continues to expose the crash being investigated.

Reducing the code that must be inspected is also part of diagnosing crashes in generated harnesses. Arash checks for API misuse and selects source statements relevant to the parameters involved in a crash before supplying them to an LLM~\cite{arash2026}. This focuses the analysis on arguments and operations that may explain the crash. FuzzAgent first minimizes the harness, then combines crash reproduction, source inspection, and debugging to distinguish library bugs from API misuse~\cite{fuzzagent2026}. HarnessMinimizer automatically removes irrelevant code from a library fuzzing harness while preserving the original crash. It embeds the input values in the resulting reproducer to help developers and LLMs inspect the calls and arguments involved in the crash.

Producing this reproducer requires comparing the crash from each candidate with the reference crash throughout minimization. A comparison based on function names and source locations first requires symbolization, which translates instruction addresses using the binaries' symbols and debug information. Repeating this step for each crashing candidate adds to reduction time even when the trace comparison itself is fast. HarnessMinimizer instead uses a normalized crash pattern together with a library-relative offset, or unsymbolized stack depth when a shared-library frame is unavailable, to filter candidates during reduction (\secref{sec:design-crash-preservation}). 

\section{Conclusion}
\label{sec:conclusion}

We presented \sysname to make automatic reduction of library fuzzing harnesses faster and the resulting reproducers easier to inspect. Its 3.46--4.51$\times$ speedups across four engines show that reusing preparation and reducing diagnostic work can benefit different reduction strategies. Input replay enables more extensive simplification by allowing removal of operations retained only to preserve consumption order. The improved identification of harness bugs in the LLM triage study shows that smaller reproducers with explicit input values can also aid crash triage. These findings support using efficient harness minimization between crash discovery and triage to help investigate library defects and harness misuse.

\section*{Tool Availability}

\sysname is publicly available at
\url{https://github.com/FuzzAnything/HarnessMinimizer}.

\bibliographystyle{ACM-Reference-Format}
\bibliography{references}

\end{document}